\documentclass[prb,reprint,a4paper,superscriptaddress,showpacs]{revtex4-1}	
\relax
\usepackage{graphicx}
\usepackage{epsfig}
\usepackage{epstopdf}
\usepackage{amsmath}
\usepackage{amssymb}
\usepackage{hyperref}
\usepackage{dcolumn}	

\begin{document}
\title{No Evidence for Orbital Loop Currents in Charge Ordered YBa$_2$Cu$_3$O$_{6+x}$ from Polarized Neutron Diffraction}	
\author{T. P. Croft}
\affiliation{H. H. Wills Physics Laboratory, University of Bristol, Bristol, BS8 1TL, United Kingdom.}

\author{E. Blackburn}
\affiliation{School of Physics \& Astronomy, University of Birmingham, Birmingham B15 2TT, United Kingdom.}
\author{J. Kulda}
\affiliation{Institut Laue-Langevin, 6, rue Jules Horowitz, BP 156, 38042 Grenoble Cedex 9, France.}
\author{Ruixing Liang}
\affiliation{Department of Physics \& Astronomy, University of British Columbia, Vancouver, Canada.}
\affiliation{Canadian Institute for Advanced Research, Toronto, Canada.}
\author{D. A. Bonn}
\affiliation{Department of Physics \& Astronomy, University of British Columbia, Vancouver, Canada.}
\affiliation{Canadian Institute for Advanced Research, Toronto, Canada.}
\author{W. N. Hardy}
\affiliation{Department of Physics \& Astronomy, University of British Columbia, Vancouver, Canada.}
\affiliation{Canadian Institute for Advanced Research, Toronto, Canada.}
\author{S. M. Hayden}
\email{s.hayden@bris.ac.uk}
\affiliation{H. H. Wills Physics Laboratory, University of Bristol, Bristol, BS8 1TL, United Kingdom.}

\begin{abstract}	
It has been proposed that the pseudogap state of underdoped cuprate superconductors may be due to a transition to a phase which has circulating currents within each unit cell. Here, we use polarized neutron diffraction to search for the corresponding orbital moments in two samples of underdoped YBa$_2$Cu$_3$O$_{6+x}$ with doping levels $p=0.104$ and 0.123. In contrast to some other reports using polarized neutrons, but in agreement with nuclear magnetic resonance and muon spin rotation measurements, we find no evidence for the appearance of magnetic order below 300\;K. Thus, our experiment suggests that such order is not an intrinsic property of high-quality cuprate superconductor single crystals. Our results provide an upper bound for a possible orbital loop moment which depends on the pattern of currents within the unit cell.  For example, for the CC-$\theta_{II}$ pattern proposed by Varma, we find that the ordered moment per current loop is less than 0.013\;$\mu_B$ for $p=0.104$.

\end{abstract}
\maketitle
\section{Introduction}	

In addition to their high transition temperatures, an ubiquitous feature of the cuprate superconductors is the existence of a normal state pseudogap (PG)\cite{Warren1989_WWBC,Ding1996_DYCT,Timusk1999_TiSt} for underdoped compositions. The pseudogap state corresponds to a loss of low-energy electronic spectral weight and has been observed by many thermodynamic and spectroscopic probes \cite{Warren1989_WWBC,Ding1996_DYCT,Timusk1999_TiSt} including nuclear magnetic resonance (NMR) and angle-resolved photo-emission spectroscopy (ARPES).
The origin of the pseudogap is not yet understood, however many believe that it holds the key to understanding the high-temperature superconductivity (HTC) phenomenon.  Specifically, it has been suggested \cite{Chakravarty2001_CLMN} that the pseudogap is due to a broken symmetry state.  The nature of the broken symmetry remains to be determined. There are many proposals \cite{Chakravarty2001_CLMN}, including: staggered fluctuating currents, loop currents which conserve translational symmetry, $d$-density waves and other possibilities.

\begin{figure*}
\label{Fig:orb_pattern_3d}
\includegraphics[width=0.95\linewidth]{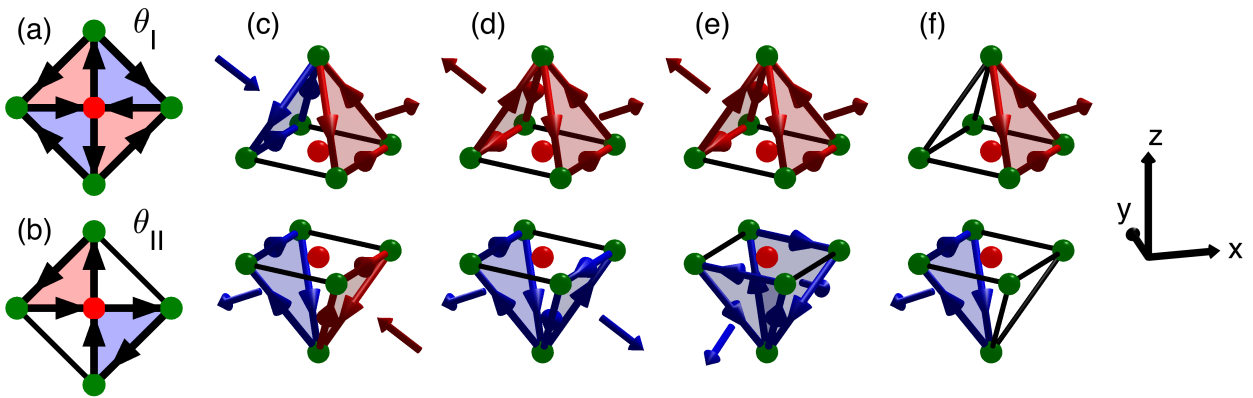}
\caption{ Some possible configurations of the CC-order in YBa$_2$Cu$_3$O$_{6+x}$. Panels (a,b) show the original CC-$\theta_{I}$ and CC-$\theta_{II}$ states proposed by Varma \cite{Varma2006_Varm} in which current flows within the CuO$_2$ planes. (c)-(f) Show arrangements based on those proposed by Yakovenko \cite{Yakovenko2015_Yako, Lederer2012_LeKi} in which current flows out of the CuO$_2$ planes. The oxygen pyramids which make up the bilayer structure are shown with copper ions as red spheres and oxygen ions as green. Arrows connecting ions indicate the direction of orbital current flow.  Arrows outside the unit cell indicate the resulting moments and are colored red for those with a component along the $z$-axis and blue for those with one along $-z$.}
\end{figure*}

This paper focusses on the predictions of a model for the cuprates proposed by Varma \cite{Varma1997_Varm,Varma2006_Varm} in which there is a continuous transition to a phase which has circulating currents (CC) within each unit cell.  The new phase preserves the translational periodicity of the crystal but breaks time reversal symmetry.  Fig.\ref{Fig:orb_pattern_3d}(a,b) show the $\theta_{I}$ and $\theta_{II}$ broken symmetry CC states for a single CuO$_2$ plane of YBa$_2$Cu$_3$O$_{6+x}$ (YBCO) originally proposed by Varma \cite{Varma2006_Varm}.  Circulating currents states lead to microscopic orbital magnetic moments which should be detected by probes such as neutron scattering, nuclear magnetic resonance and muon spin rotation.  The experimental evidence with regard to the existence of these moments is unclear.   Early spin-polarized neutron scattering measurements \cite{Lee1999_LMSS} on YBa$_2$Cu$_3$O$_{6+x}$ and La$_{2-x}$Sr$_x$CuO$_4$  failed to observe a magnetic moment due to the $\theta_{I}$ state.  However, later polarized neutron studies on underdoped YBa$_2$Cu$_3$O$_{6+x}$ (YBCO) reported \cite{Fauque2006_FSHP,Mook2008_MSFB,Bourges2011_BoSi,Sidis2013_SiBo,Mangin-Thro2015_MSWB,Mangin-Thro2017_MLSB} a $\mathbf{q}=0$ magnetic order with long range 3D correlations \footnote{The width of the peak in Fig. 2(a) of {Fauqu\'{e}} et al. \cite{Fauque2006_FSHP} and Mook et al. \cite{Mook2008_MSFB} give lower bounds for {$\xi_c$} of 130~\AA\ and 75~\AA\ respectively for samples with {$p=0.091$} amd {$p=0.112$}.}, which can be interpreted as an evidence for other states (like CC-$\theta_{II}$) with intra-unit cell circulating currents corresponding to moments of $\sim$0.1\;$\mu_B$.  There are also reports of moments being observed in HgBa$_2$CuO$_{4+x}$ (Hg1201) \cite{Li2008_LBBC}, Bi$_2$Sr$_2$CaCu$_2$O$_{8+x}$ (Bi2212)\cite{DeAlmeida-Didry2012_DSBG} and La$_{2-x}$Sr$_{x}$CuO$_{4}$ (LSCO)(short-ranged) \cite{Baledent2010_BFSC}.

Orbital current order with moments of $\sim$0.1$-$0.2\;$\mu_B$ per triangle suggested by some neutron scattering measurements \cite{Bourges2011_BoSi} should also be detectable \cite{Lederer2012_LeKi} by NMR measurements.  However NMR measurements \cite{Wu2011_WMKH,Wu2015_WMKH,Straessle2011_SGMR,Kawasaki2010_KLKR,Mounce2013_MOLH} on
YBa$_2$Cu$_3$O$_{6+x}$, YBa$_2$Cu$_4$O$_8$, Bi$_2$Sr$_{2-x}$La$_x$CuO$_{6+x}$ and HgBa$_2$CuO$_{4+x}$ found no evidence of orbital order. In addition, muon spin rotation measurements ($\mu$SR) on YBa$_2$Cu$_3$O$_{6+x}$  and La$_{2-x}$Sr$_x$CuO$_4$ did not detect magnetic order \cite{Sonier2001_SBKM,Sonier2002_SBKH,MacDougall2008_MACI,Pal2016_PAPI} of the strength $\sim 0.1$\;$\mu_B$ suggested by Refs.~\onlinecite{Fauque2006_FSHP, Mook2008_MSFB} or they detect signals \cite{Sonier2009_SPSH} which do not correlate with the neutron scattering measurements \cite{Fauque2006_FSHP, Mook2008_MSFB}.  It is possible that the magnetic moments are fluctuating just slowly enough to appear static to neutrons, but too fast to be identified as magnetic order by NMR \cite{Wu2015_WMKH} and $\mu$SR \cite{Pal2016_PAPI}. However, a magnetic phase transition requires a finite order parameter or finite time-averaged moments, and in any case, our present measurements are in agreement with NMR and $\mu$SR.

Another probe which can detect time reversal symmetry breaking (TRSB) is the polar Kerr effect (PKE). High resolution measurements  \cite{Xia2008_XSDK} have detected a PKE effect below a temperature $T_{\textrm{Kerr}}$ in YBa$_2$Cu$_3$O$_{6+x}$. Fig.~\ref{YBCO_phase_diagram} shows a comparison of $T_{\textrm{Kerr}}$ with onset temperature $T_{\mathrm{mag}}$ of the $\mathbf{q}=0$ magnetic order detected by polarized neutrons in Refs.~\onlinecite{Fauque2006_FSHP, Mook2008_MSFB}. The onset temperatures do not agree suggesting that the two probes may be observing different phenomena.

Charge density wave (CDW) order has recently been observed in underdoped YBa$_2$Cu$_3$O$_{6+x}$ by NMR \cite{Wu2011_WMKH} and x-ray diffraction \cite{Ghiringhelli2012_GLMB,Chang2012_CBHC}.  This has added additional complexity to the phase diagram of YBCO (see Fig.~\ref{YBCO_phase_diagram}).  The CDW competes with superconductivity\cite{Ghiringhelli2012_GLMB,Chang2012_CBHC} and it is natural to ask whether it also competes with the reported CC order observed using polarized neutrons \cite{Fauque2006_FSHP,Mook2008_MSFB,Bourges2011_BoSi,Sidis2013_SiBo,Mangin-Thro2015_MSWB,Mangin-Thro2017_MLSB}.   This motivated us to investigate the CC-order with polarized neutrons in samples in which the CDW had been observed \cite{Chang2012_CBHC,Blackburn2013_BCHH,Hucker2014_HCHB} by x-rays.   We use high-quality detwinned samples with a mosaic spread less than 0.1$^{\circ}$ grown by a self-flux method \cite{Liang1998_LiBH}. Although our crystals are approximately two orders of magnitude smaller than those used by Fauqu\'{e} \textit{et al.} \cite{Fauque2006_FSHP} in their initial report on possible CC order, our experiment has the required sensitivity. This can be seen from the errors quoted in our final experimental results and in the figures. These errors derive from the number of neutrons counted. We observed no evidence of the previously reported\cite{Fauque2006_FSHP, Mook2008_MSFB} $\mathbf{q}=0$ magnetic order.  It should be noted that the present experiment was carried out on an instrument with a factor of $\sim$3 times higher flux and up to $\sim$10 times longer counting times are used at each temperature.  

\section{Background}
\label{Sec:background}

\subsection{Neutron Cross Sections}
\label{Sec:neutron_cross_sec}
Neutrons interacting with matter are scattered both by atomic nuclei (nuclear strong interaction) and by the orbital and spin magnetic moments of the electrons (electromagnetic interaction). Bragg scattering occurs when the neutron momentum transfer (scattering vector) $\mathbf{Q}= \mathbf{k}_i - \mathbf{k}_f$ equals a reciprocal lattice vector $\mathbf{G}$.  The resulting nuclear Bragg peaks reflect the chemical crystal structure in the first case.  In the second case the magnetic Bragg peaks, appearing at the same or different reciprocal space positions as the  nuclear ones, provide information about the magnetic order. Only the component of the local magnetization density $\textbf{M}(\mathbf{r})$ perpendicular to the momentum transfer (scattering vector), $\textbf{M}_{\perp}(\mathbf{r})=\textbf{M}(\mathbf{r})-(\textbf{M}(\mathbf{r})\cdot \hat{\textbf{Q}})\hat{\textbf{Q}}$, contributes to the scattering cross section. This component can be further split into parts perpendicular and parallel to the neutron spin direction, giving rise to partial cross sections corresponding to scattering processes inverting (in the first case) and conserving (in the second case) the neutron spin orientation. Usually they are referred to as spin-flip (SF, $\uparrow\downarrow$)  and non spin-flip (NSF, $\uparrow\uparrow$) processes.

Neutron polarization analysis \cite{Williams1988_Will, Shirane2002_ShSM, Moon1969_MoRK, Squires1996_Squi} 
in the neutron scattering experiments may be used to detect even a small magnetic contribution in the presence of a strong nuclear  Bragg intensity by using selection rules specific to the neutron spin behavior in the magnetic scattering process.  Its simplest implementation - the longitudinal polarization analysis - consists in preparing a beam with one neutron spin orientation. Neutrons are then scattered under a small guiding field of the order of 1.5 mT and the number of neutrons in each final spin state is measured. 

In the present experiment the guide field is used to align the neutron polarization $\mathbf{P}$ parallel to $\mathbf{Q}$.  The SF and NSF cross-sections are then given by \cite{Moon1969_MoRK, Squires1996_Squi},

\begin{eqnarray}
\sigma_{\uparrow\downarrow}=\left( \frac{d \sigma}{d \Omega}\right)_{\uparrow\downarrow}^{\mathbf{P} \parallel \mathbf{Q}}   &=&  \left( \frac{\gamma r_0}{2 \mu_B} \right)^2 \left|\mathbf{M}_{\perp}(\mathbf{G}) \right|^2 , 
\label{Eqn:xsct_mag} \\
\sigma_{\uparrow\uparrow}=\left( \frac{d \sigma}{d \Omega}\right)_{\uparrow\uparrow}^{\mathbf{P} \parallel \mathbf{Q}} &=&  \left| F_N (\textbf{G}) \right|^2, 
\label{Eqn:xsctn_nuc}
\label{Eqn:xsct_nmag}
\end{eqnarray}
where the Fourier component of the local magnetization density $\textbf{M}(\mathbf{r})$ is given by
\begin{equation}
\mathbf{M}(\mathbf{G})=\int_{\mbox{unit cell}} \mathbf{M}(\mathbf{r}) \exp(i \mathbf{G} \cdot \mathbf{r}) \, d\mathbf{r},
\label{Eqn:SF_mag}
\end{equation}
$(\gamma r_0/2)^2 = 7.18 \times 10^{-30}\;$m$^2$=71.8\;mbarn, $F_N (\textbf{G})$ is the nuclear structure factor \cite{Squires1996_Squi} and
\begin{equation}
\textbf{M}_{\perp}(\mathbf{G})=\textbf{M}(\mathbf{G})-(\textbf{M}(\mathbf{G})\cdot \hat{\textbf{G}})\hat{\textbf{G}}.
\label{Eqn:M_perp}
\end{equation}
We refer to $\textbf{M}_{\perp}(\mathbf{G})$ as the magnetic structure factor since from Eq.~\ref{Eqn:xsct_mag} the magnetic scattering is proportional to its modulus squared. 

\subsection{Calculation of the scattering structure factor of the orbital current patterns}

A number of CC-states have been proposed to explain the PG  \cite{Chakravarty2001_CLMN,Varma1997_Varm,Simon2002_SiVa,Varma2006_Varm, Lederer2012_LeKi,Yakovenko2015_Yako, Mangin-Thro2017_MLSB}. A selection of states which break time reversal symmetry while preserving lattice translational symmetry are shown in Fig.~\ref{Fig:orb_pattern_3d}. The original model of Varma \cite{Varma2006_Varm} considers a single CuO$_2$ layer. YBCO has a bilayer structure formed by CuO$_5$ pyramids. The presence of in-plane loop currents may lead to out-of-plane loop currents\cite{Lederer2012_LeKi,Yakovenko2015_Yako} involving the apical oxygens in the YBCO structure. Hence, we also consider some of these patterns. All the CC-$\theta_{I}$ and CC-$\theta_{II}$ broken symmetry states have the translation symmetry of the CuO$_2$ lattice and therefore do not induce Bragg scattering at new reciprocal lattice positions. All these states are examples of $\mathbf{q}=0$ antiferromagnetic (AFM) order, i.e. magnetic order where the ordering pattern is the same in all crystallographic unit cells.   

The magnetic structure factors $\left|\mathbf{M}_{\perp}(\mathbf{G}) \right|$ corresponding to models (a)-(f) and for various reciprocal lattice positions were computed numerically from Eqs.~\ref{Eqn:SF_mag}--\ref{Eqn:M_perp} and displayed in Table~\ref{Table:M_perp_cals}. We assume the moment is spread uniformly over the shaded triangles in Fig.~\ref{Fig:orb_pattern_3d} and arises from current flowing between the centers of the copper and oxygen ions, that is we do not explicitly take into account the atomic orbitals. The magnetic moment associated with each triangular loop is denoted as $\mathbf{m}_0$. In the case of (a) and (b) we assume the moment pattern is the same in the two CuO$_2$ planes of the bilayer.  We regard the patterns shown in Fig.~\ref{Fig:orb_pattern_3d} as representative for the purposes of interpreting the present experiment which requires a specific model in order to convert a measured cross-section into a microscopic moment.

\begin{table}[t]

\begin{ruledtabular}
\begin{tabular}{c|rrrrrr}
 $(hkl)$ & \multicolumn{6}{c}{$\left|\mathbf{M}_{\perp}(\mathbf{G}) \right|/\left| \mathbf{m}_0 \right|$} \\ \hline
      & (a) & (b)  & (c)  & (d)  & (e)  & (f)  \\
(100)   & 0.00	& 2.55	& 1.16	& 1.60	& 0.00	& 0.99 \\
(010)	& 0.00	& 2.55	& 1.16	& 1.60	& 0.00	& 0.99 \\
(110)	& 3.24	& 0.00	& 0.00	& 0.00	& 0.74	& 0.00 \\
(011)	& 0.00	& 2.36	& 0.33	& 0.39	& 1.04	& 0.35 \\
(012)	& 0.00	& 1.91	& 0.21	& 0.41	& 0.19	& 0.26 \\
(020)	& 0.00	& 1.27	& 0.58	& 0.80	& 0.00	& 0.49 \\
\end{tabular}
\end{ruledtabular}
\caption{ \label{Table:M_perp_cals} Calculated values of the magnetic structure factor for the orbital current patterns in Fig.~\ref{Fig:orb_pattern_3d}. The left most column denotes the reciprocal lattice position $\mathbf{G}=h \mathbf{a}^{\star}+k \mathbf{b}^{\star}+l \mathbf{c}^{\star}$. Throughout the paper, we use the $Pmmm$ space group and the unit cell with $a \approx 3.84$\;\AA, $b\approx 3.88$\;\AA\ and $c \approx 11.7$\;\AA. The column headers (a)--(f) refer to the patterns in Fig.~\ref{Fig:orb_pattern_3d}.  $\left|\mathbf{M}_{\perp}(\mathbf{G}) \right|$ is calculated using Eqn.~\ref{Eqn:SF_mag} for a formula unit (f.u.) of YBa$_{2}$Cu$_3$O$_{6+x}$ and $\left| \mathbf{m}_0 \right|$ is the moment of a single triangular loop of orbital current.
}
\end{table}

\begin{figure}
\includegraphics[width=0.9\linewidth]{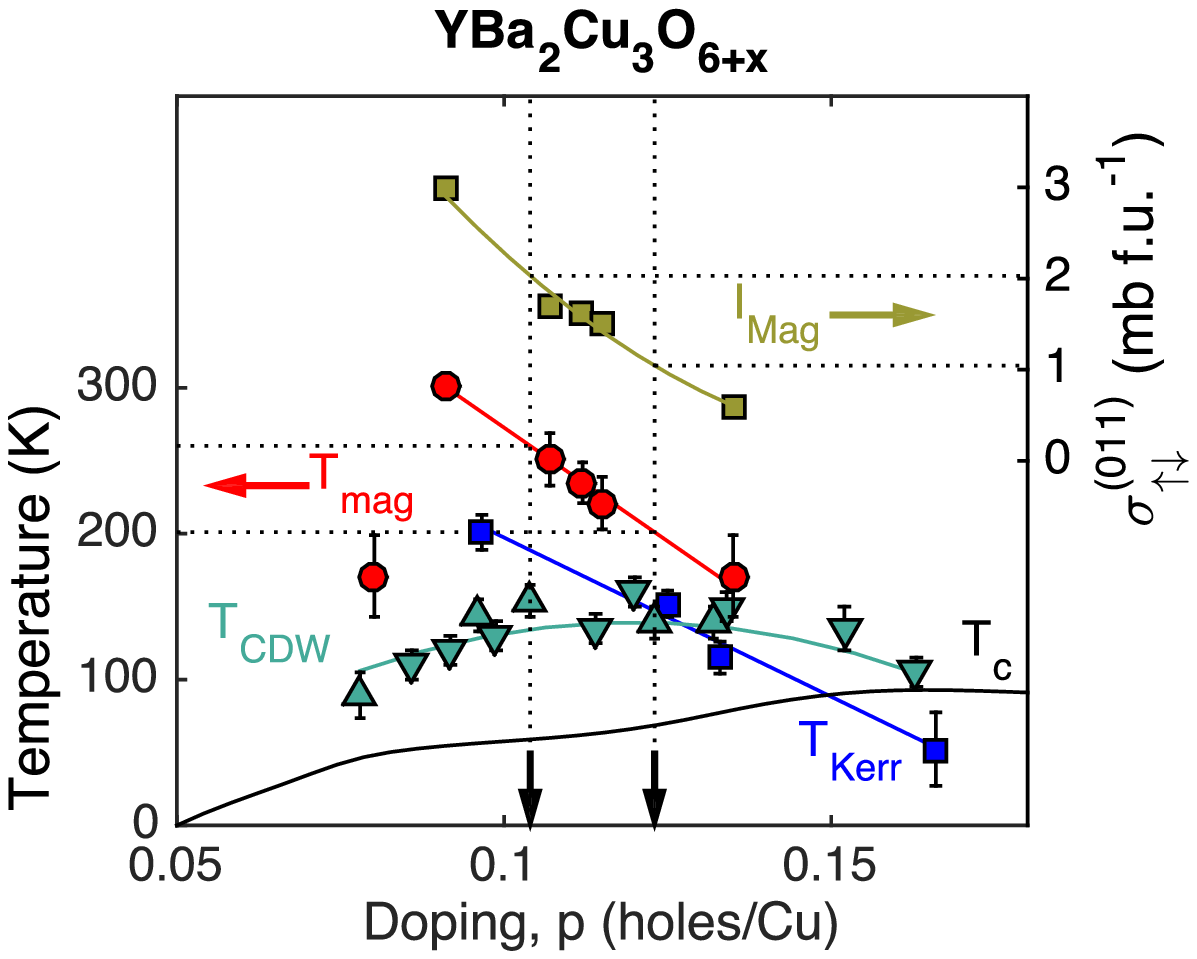}
\caption{\label{YBCO_phase_diagram} YBCO phase diagram. (left axis) Superconducting transition temperature, $T_c$, versus hole doping $p$ (black line).  Also included are onset temperatures of CDW order $T_\mathrm{CDW}$ ($\blacktriangle$, \onlinecite{Hucker2014_HCHB} and $\blacktriangledown$, Ref.~\onlinecite{Blanco-Canosa2014_BFSP}), putative $\mathbf{q}=0$ AFM order $T_\mathrm{mag}$~\cite{Fauque2006_FSHP,Mook2008_MSFB,Baledent2011_BHSH} and the Kerr anomaly $T_\mathrm{Kerr}$~\cite{Xia2008_XSDK}. (right axis) Doping dependence of the magnetic intensity at the (011) Bragg refection reported in Refs.~\onlinecite{Fauque2006_FSHP,Mook2008_MSFB,Baledent2011_BHSH,Bourges2017_BoSM}. Doping levels of the $x=0.54$ and $x=0.67$ samples used in the present study are denoted by the arrows and the horizontal dotted lines indicate corresponding values of $T_\mathrm{mag}$ and $\sigma_{\uparrow\downarrow}^{(011)}(T \approx T_c)$.}
\end{figure}

\subsection{Polarized Beam Experiments}

The observed neutron counting rate is related to the cross-section $\sigma$ by $I = I_{0} V \sigma + I_{BG}$, where $I_{0}$ and $V$ denote  a general scale factor (containing the incident neutron flux) and the sample volume and $I_{BG}$ is a background, potentially different for each particular spin orientation. It is common practice in polarized neutron experiments of this type, to treat data in terms of a flipping ratio $R$, which has the advantage that all the multiplicative terms entering the scale factor cancel out and we may hope to obtain directly the ratio of the two corresponding cross-sections. Thus, for an ideal measurement (no background, completely polarized beam and perfect neutron spin analysis) we would have,
\begin{equation}
\frac{\sigma_{\uparrow\uparrow}}{\sigma_{\uparrow\downarrow}}=R,
\label{Eqn:FR}
\end{equation}
and,
\begin{equation}
\sigma_{\uparrow\downarrow}  =\left( \frac{\gamma r_0}{2 \mu_B} \right)^2 \left|\mathbf{M}_{\perp}(\mathbf{G}) \right|^2 = |F_N (\textbf{G})|^2 \times R^{-1}.
\label{Eqn:sigma_R}
\end{equation}
In practice, we use the background-corrected intensities measured for two different neutron spin orientations at otherwise unchanged experimental conditions $R_{\textrm{meas}}=\left(I_{NSF} - I_{BG}\right)/\left(I_{SF} -I_{BG}\right)$. 

A real instrument has imperfections such as the finite efficiencies of the polarizer and analyzer. This leads to a mixing of the two measured cross sections. This mixing may be treated by taking into account the effective beam polarization $P$ ($P < 1$) such that a beam fraction $P$ is sensitive to the cross-section of interest ($\sigma_{\uparrow\uparrow}$ or $\sigma_{\uparrow\downarrow}$) and a fraction $(1-P)$ equally sensitive to both cross-sections ($\sigma_{\uparrow\uparrow}$ and $\sigma_{\uparrow\downarrow}$).  This corresponds to an instrumental flipping ratio $R_{\mathrm{inst}}=\left(1+P\right)/\left(1-P\right)$. For a real instrument, the measured flipping ratio corresponds then to a mixture of cross-sections:
\begin{equation}
R_{\mathrm{meas}}=\frac{P\sigma_{\uparrow\uparrow}+\left(1-P\right)\left(\sigma_{\uparrow\uparrow}+\sigma_{\uparrow\downarrow}\right)/2}{P\sigma_{\uparrow\downarrow}+\left(1-P\right) \left(\sigma_{\uparrow\uparrow}+\sigma_{\uparrow\downarrow}\right)/2}.
\label{Eqn:rmeas}
\end{equation}
If we can neglect $\sigma_{\uparrow\downarrow}$ with respect to $R_{\mathrm{inst}}\sigma_{\uparrow\uparrow}$ the above equation can be simplified to 
\begin{equation}
\sigma_{\uparrow\downarrow} = \sigma_{\uparrow\uparrow} \left[\frac{1}{R_{\mathrm{meas}}} -\frac{1}{R_{\mathrm{inst}}} \right],  
\label{Eqn:sigma_est}
\end{equation}
which will serve as a fundamental reference for our experiment.

Combining Eqn.~\ref{Eqn:sigma_R} and \ref{Eqn:sigma_est}, we obtain an estimate for the magnetic structure factor from the measured flipping ratio\cite{Fauque2006_FSHP}:
\begin{equation}
\left| \mathbf{M}_{\perp}(\mathbf{G}) \right|^2 = \left( \frac{2 \mu_B}{\gamma r_0} \right)^2 
\left| F_N (\textbf{G}) \right|^2 \left[ \frac{1}{R_{\mathrm{meas}}} - \frac{1}{R_{\mathrm{inst}}} \right]. 
\label{Eqn:FR2M_perp}
\end{equation}
A similar equation was used by Fauqu\'{e} \textit{et al.} \cite{Fauque2006_FSHP}.

\begin{figure}
\includegraphics[width=1.0\linewidth]{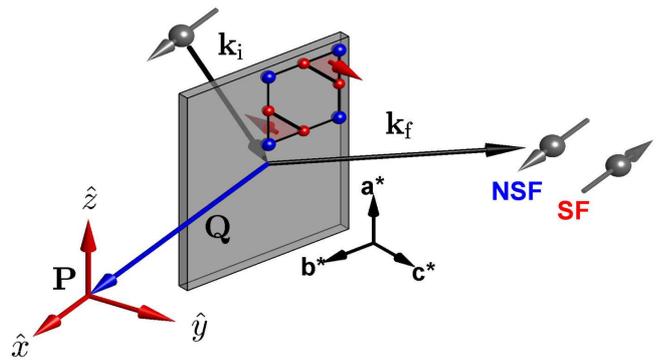}
\caption{\label{scattgeom} Longitudinal polarization analysis (LPA) setup used in this experiment. The neutron polarization used was $\textbf{P}\parallel\textbf{Q}$. The initial and final wavevectors of the neutron are labelled by $\mathbf{k}_i$ and $\mathbf{k}_f$.}
\end{figure}

\section{Experimental Method}
\label{exp_method}
\subsection{Sample Details}
We have investigated two samples with compositions YBa$_2$Cu$_3$O$_{6.54}$ and YBa$_2$Cu$_3$O$_{6.67}$, further details are given in Table~\ref{Tbl:samplesum}.   Each of our samples consisted of one high-quality single crystal detwinned to $\sim99 \%$\cite{Chang2012_CBHC}.  The YBa$_2$Cu$_3$O$_{6.67}$ sample had dimensions 3.1 $\times$ 1.7 $\times$ 0.6 mm$^3$ and a mass 18~mg (see inset of Fig.~\ref{orthoVIII_raw_IFR}); the other sample was similar. Both samples where prepared by the self flux method using a BaZrO$_3$ crucibles\cite{Erb1995_ErWF,Liang1998_LiBH}.  Samples grown by the same method and group \cite{Liang1998_LiBH} have an overall purity of greater than 99.99 \%.  Polarized optical microscopy reveals no evidence of secondary phases.  Our crystals have been characterized by x-ray diffraction and both do exhibit charge density waves\cite{Chang2012_CBHC,Chang2016_CBIH,Blackburn2013_BCHH}. Hard (100 keV) x-ray diffraction\cite{Chang2012_CBHC} on the (020) reflection [see Fig.~\ref{orthoII_020FR_65K}(a)] indicates an overall bulk mosaic spread of less than 0.1$^{\circ}$ for each of the crystals. It should be noted that these x-ray measurements are performed in transmission so that the bulk (rather than surface) of the sample is probed.  Another indication of the high crystalline quality (electronic mean free path) is that samples prepared by the same method and growers exhibit quantum oscillations \cite{Doiron-Leyraud2007_DPLL}. Our samples contrast with those used by other groups \cite{Fauque2006_FSHP, Mook2008_MSFB,Mangin-Thro2017_MLSB} which have larger areas/volumes and broader mosaic distributions.  The experiments of Fauqu\'{e} \textit{et al.} \cite{Fauque2006_FSHP} and Mangin-Thro \textit{et al.} \cite{Mangin-Thro2017_MLSB} were carried out on arrays of self-flux grown samples with overall mosaics in the range 1.2--2.2$^{\circ}$ while Mook \textit{et al.} \cite{Mook2008_MSFB} investigated a melt-processed sample \cite{Murakami1992_Mura} with a mass of 25g.    

\begin{table}[b]
\begin{ruledtabular}
\begin{tabular}{lllllll}
$y$ in & O & $p$ & $T_\mathrm{c}$ & $T_\mathrm{CDW}$ & $T_{\mathrm{mag}}$ & $\sigma_{\uparrow\downarrow}(011)$ \\
YBCO & order &  & (K) & (K) & (K) & (mb\;f.u.$^{-1}$) \\ \hline
6.54 & o-II & 0.104 & 58 & 155(10) & 259 & 2.0\\
6.67 & o-VIII & 0.123 & 67 & 140(10) & 201 & 1.0 \\
\end{tabular}
\end{ruledtabular}
\caption{\label{Tbl:samplesum} Properties of the two YBCO samples studied. Planar doping $p$ was determined as in Liang \textit{et al.}\cite{Liang2006_LiBH}. $T_\mathrm{c}$ was determined by 1 Oe field-cooled magnetization. The onset of CDW order, $T_\mathrm{CDW}$, was identified using hard x-rays in Ref.~\onlinecite{Blackburn2013_BCHH}. $T_\mathrm{mag}$ are the estimated (see Fig.~\ref{YBCO_phase_diagram}) onset temperature of the putitive $\mathbf{q}=0$ magnetic order observed in Fauqu\'{e} \textit{et al.} and related papers \cite{Fauque2006_FSHP,Mook2008_MSFB,Baledent2011_BHSH,Bourges2017_BoSM}. $\sigma_{\uparrow\downarrow}(011)$ is the
 corresponding intensity of the magnetic signal for the (011) Bragg position for $T \approx T_c$.}
\end{table}

\subsection{Polarized neutron diffraction}

\begin{figure}
\includegraphics[width=1.0\linewidth]{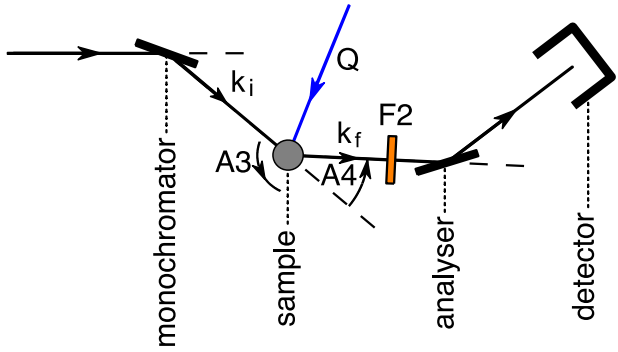}
\caption{\label{IN20schem} Schematic of the IN20 spectrometer in the horizontal plane. Neutrons are monochromated and polarized using Heusler-alloy monochromator. The neutron spin-state is maintained using a small guide field throughout the spectrometer. The Heusler analyzer scatters spin of one polarization to the detector. A Mezei coil flipper \cite{Williams1988_Will} F2 can be switched on to flip the polarization of neutrons. This determines which polarization the analyzer system detects. }
\end{figure}

In order to obtain the high flux and instrumental flexibility required for our measurement we used a triple-axis spectrometer with polarization analysis \cite{Shirane2002_ShSM, Note_IN20, Williams1988_Will} similarly to other groups\cite{Lee1999_LMSS,Fauque2006_FSHP,Mook2008_MSFB,Bourges2011_BoSi,Sidis2013_SiBo,Li2008_LBBC,DeAlmeida-Didry2012_DSBG,Baledent2010_BFSC} searching for orbital loop currents. The present experiment was performed on the IN20 triple-axis spectrometer (see Fig.~\ref{IN20schem}) at the Institut Laue-Langvin (ILL) in Grenoble, France. IN20 has the excellent angular positioning capability required for the experiment with the rotations A3 and A4 (Fig.~\ref{IN20schem}) being reproduced to 0.01$^{\circ}$ and 0.02$^{\circ}$ respectively. Our samples were mounted with the $\textbf{a}$-axis vertical allowing access to reflections of the type $(0kl)$ in the horizontal scattering plane.  The measurements were performed with an incident neutron energy $E_i=13.6$~meV corresponding to $k_i=2.662$~\AA$^{-1}$ with a pyrolytic graphite filter placed before the sample to filter out higher order neutrons.  The incident beam was polarized and the scattered beam analyzed using Cu$_2$MnAl Heusler (111) crystals providing an overall polarization $P\approx 0.95$ (95\%), corresponding to instrumental flipping ratios $R_{\textrm{inst}}\approx 40$.  The sample position was surrounded by a coil system which allowed a small guide field ($\approx 1.5$ mT) to be applied at the sample position.  In the present experiment, this field was used to align the neutron polarization $\mathbf{P}$ parallel to the scattering vector $\mathbf{G}$ (cf. Fig. \ref{scattgeom}).  No collimators were placed between monochromator and sample or between sample and analyzer.

\subsection{Data collection method}
\label{data_collection_method}

\begin{figure}
\includegraphics[width=1.0\linewidth]{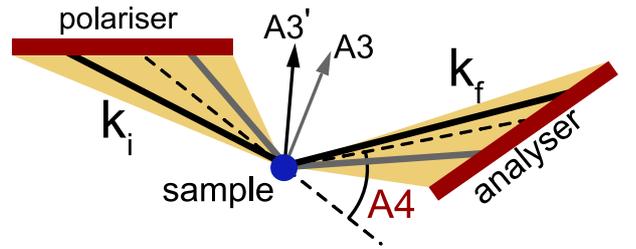}
\caption{The small mosaic of the sample means that only part of the monochromator and analyzer crystals are selected. For example, the grey (black) lines show $\mathbf{k}_i$ and $\mathbf{k}_f$ for sample rotation angle A3 (A3$^{\prime}$). At each temperature, the sample rotation (A3) and the position of the analyzer (A4) were scanned to select the same part of the monochromator and analyzer. A4 is the scattering angle corresponding to the center of analyzer.}
\label{polarizer_analyser}
\end{figure}

The polarized neutron beam arriving at the sample position has approximate horizontal and vertical divergences of 40$^\prime$  and 120$^{\prime}$, respectively, determined by the size, curvature and mosaic spread of the crystal monochromator and other factors. This large divergence is ideal for inelastic scattering studies of spin excitations in the conventional operation of IN20. In the present study we are, however, dealing with Bragg diffraction from mm-sized single crystals with mosaic spreads of less than 0.1$^{\circ}$, as determined by 100 keV x-ray scattering\cite{Chang2012_CBHC,Blackburn2013_BCHH}. In this case, Bragg's law, in combination with the small sample mosaic and the small sample size, places stringent conditions on the beam trajectory in the horizontal plane (see Fig.~\ref{polarizer_analyser}). The Bragg condition is invariant to rotation of $\mathbf{k}_i$ and $\mathbf{k}_f$ around the scattering vector $\mathbf{G}$. This means that it is much less sensitive to angular deviations of $\mathbf{k}_i$ and $\mathbf{k}_f$ in the vertical plane. As a consequence, the acceptance angles of the sample Bragg reflections subtend only narrow vertical stripes (a few mm wide and several cm tall) on the surface of the monochromator and analyzer. These consist of crystal arrays with dimensions of $\approx 200 \times 100$\,mm$^2$ (horizontal $\times$ vertical) whose whole surface is active in inelastic scattering experiments.  In the present experiment, the small active areas on the monochromator/analyzer surface can be horizontally displaced by minor changes in sample orientation and lattice parameter (cf. Fig.~\ref{polarizer_analyser}). The local orientation and mosaic spread inhomogeneities of the large crystal arrays may then lead to instrumental polarization variations of up to a few percent, which would be averaged out in inelastic scattering experiments, but which become important in our case when looking for a very weak $\sigma_{\uparrow\downarrow}$ in the presence of a strong $\sigma_{\uparrow\uparrow}$. In an early search for orbital loop currents, Lee et al.\cite{Lee1999_LMSS} showed how flipping ratios can vary with sample rotation because of this.

\begin{figure}
\includegraphics[width=1.0\linewidth]{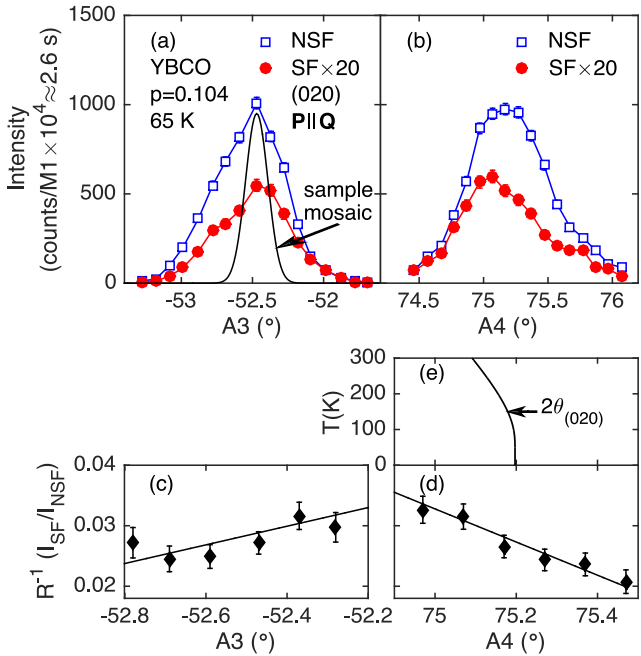}
\caption{(a) Rocking curves of a strong, exclusively nuclear Bragg peak (020) as a function of A3 and A4. The FWHM widths are 0.58$^{\circ}$ and 0.67$^{\circ}$ respectively. (b) Data in the spin-flip (SF) channel has been multiplied by a factor of 20 for presentation. (c)-(d) Variation of the corresponding inverse flipping ratio $R_{\textrm{inst}}^{-1}$  over the rocking curve ranges of panes (a)-(b).  Data were collected at 65K. The solid line in (a) indicates the sample mosaic profile as measured by 100\;keV x-ray diffraction\cite{Chang2012_CBHC}.  (e) The $T$-dependence of the scattering angle $2 \theta$ of the (020) Bragg reflection based on the data of Bo\v{z}in \textit{et al.} \cite{Bozin2016_BHSC}. } 
\label{orthoII_020FR_65K}
\end{figure}

Fig.~\ref{orthoII_020FR_65K}(a,b) show spin-flip (SF) and non-spin-flip (NSF) scans through the (020) Bragg peak as a function of instrument angles A3 (the sample rotation) and A4 (the position of the analyzer). By varying the A3 and A4 angles these scans progressively sample different parts of the monochromator and analyzer and hence we might expect the $R_{\mathrm{inst}}$ to vary with them.  Fig.~\ref{orthoII_020FR_65K}(c,d) show that this is indeed the case,  $R_{\mathrm{inst}}^{-1}$ varies with A3 and A4. Since this quantity directly affects the extraction of our magnetization signal (Eqn.~\ref{Eqn:FR2M_perp}), we minimized this effect by making sure that as the temperature is varied in our experiments, neutrons always emanate from the same part of the monochromator and strike the same part of the analyzer. Fig.~\ref{orthoII_020FR_65K}(e) shows the expected temperature variation \cite{Bozin2016_BHSC} the scattering angle $2 \theta$ of the (020) Bragg refection due to the thermal expansion of the sample (similar changes are expected for the (010) and (011) reflections).  These changes are $\sim 0.1^{\circ}$ over the temperature range investigated and would lead to different parts of the monochromator and analyzer being sampled if A3 and A4 were not moved to compensate for the change in $2 \theta$.  It was found to be unnecessary to compensate for small changes in sample height ($\sim$1\;mm) due to thermal contraction of the sample mounting stick. This is consistent with the vertically extended (several cm) beam spots on the monochromator and analyzer, mentioned at the beginning of this section.

\begin{figure}
\includegraphics[width=1.0\linewidth]{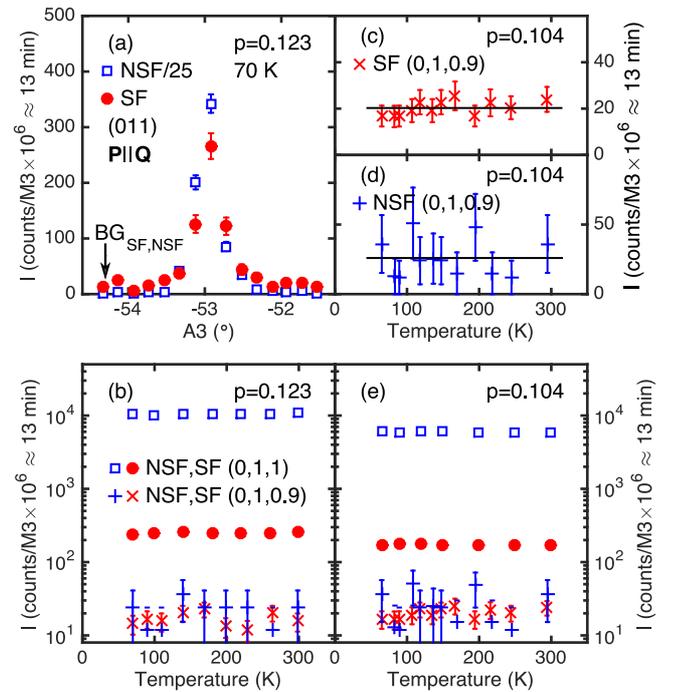}

\caption{(a) Spin-flip (SF) and non spin-flip (NSF) rocking curves of the (011) Bragg reflection to illustrate the relative signal and background intensities. Scans with a smaller A3 step yield a more accurate estimate of the peak centre and yield a FWHM NSF width of 0.57$^\circ$.  (b)-(e) The SF and NSF (011) rocking curve peak intensities and the background (A3 offset by 2$^{\circ}$) as a function of temperature on linear and logarithmic scales. }
\label{bgnd_signal_comp}
\end{figure}

For a series of measurements on a particular Bragg peak such as $\textbf{G}=$(011), the following protocol was adopted for each temperature. (1) After thermal equilibrium was reached the (200) and (006) nuclear Bragg peaks were measured and aligned to the horizontal plane. (2) The spectrometer was then moved to the position of the Bragg reflection $\mathbf{Q}=\textbf{G}$ and  the NSF intensity was maximized with respect to A3 and A4 in an iterative manner. Fig.~\ref{orthoII_020FR_65K}(a,b) show examples of A3 and A4 scans used to locate the maximum (final scans were performed with 0.05$^{\circ}$ steps).  (3) With the NSF intensity maximized with respect to A3 and A4, we have alternated counts with the flipper on and off to determine the flipping ratio $R_{\textrm{meas}}$. Counting was split into segments of no more than 11 minutes with acquisition times optimized to achieve similar statistical accuracy in both of the SF and NSF channels. (4) A3 was then displaced by 2$^{\circ}$ and the background intensities BG$_{SF}$ and BG$_{NSF}$ corresponding to $\textbf{G}$ were measured. A typical outcome of this protocol is illustrated by data taken at the (011) Bragg reflection, displayed in Fig.~\ref{bgnd_signal_comp}(a). Note that the background intensities are very small.  The values of the SF, NSF and background  intensities for different temperatures are plotted in the panes (b)-(e) of Fig.~\ref{bgnd_signal_comp}, none of them exhibits a significant temperature variation.

\section{Results}
\begin{figure*}
\includegraphics[width=0.85\linewidth]{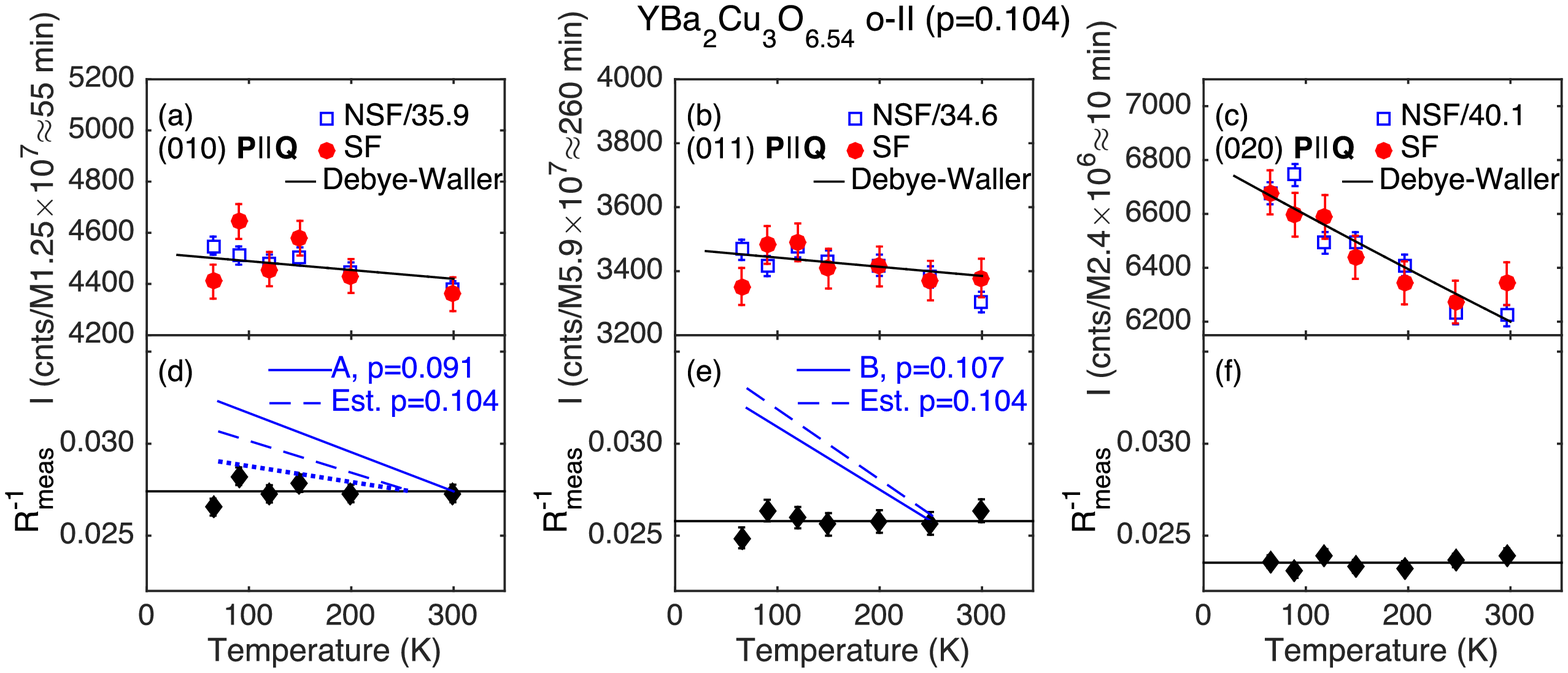}
\caption{$T$-dependence of the scattering for YBCO $x=0.54$. (a)-(c) show the raw SF/NSF scattering intensity of the (010), (011) and (020) Bragg reflections. No background subtractions have been made, the NSF data are scaled by the constants indicated. (d-f) display the inverse flipping ratios $R^{-1}_{\textrm{meas}}$ determined from the data in (a-c). In this case, a measured $T$-independent background is subtracted (see main text). Dashed lines in (d) and (e) show the estimated variation of $R^{-1}_{\textrm{meas}}$ for $p=0.104$ determined by interpolating between samples measured by Fauqu\'{e} \textit{et al.}~\cite{Fauque2006_FSHP, Bourges2017_BoSM}. Solid lines are samples with the closest doping [i.e. sample A from Fig.~2(b) and sample B from Fig.~1(c) of Ref.~\onlinecite{Fauque2006_FSHP}]. We use Eqn.~\ref{Eqn:sigma_est} with $\sigma_{\uparrow\uparrow}=|F_N|^2$=1.84 and 0.28\;barn\;f.u.$^{-1}$ for (010) and (011) respectively. See main text for more details and dotted line.}
\label{orthoII_raw_IFR}
\end{figure*}

\begin{figure}
\includegraphics[width=1.0\linewidth]{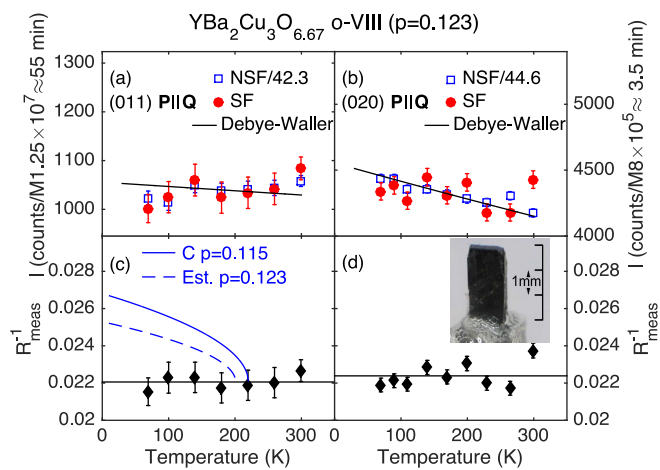}
\caption{\label{orthoVIII_raw_IFR} $T$-dependence of the scattering for YBCO $x=0.67$. Panels (a-b) show the raw scattering intensity of the SF and NSF channels of the (011) and (020) Bragg reflections. Experimental conditions and data treatment as in Fig.~\ref{orthoII_raw_IFR}. (c-d) The inverse of our measured flipping ratios $R^{-1}_{\textrm{meas}}$.  The dashed line in (c) shows the expected variation of $R^{-1}_{\textrm{meas}}$ for a $p=0.123$ sample determined by interpolating between samples measured by Fauqu\'{e} \textit{et al.}~\cite{Fauque2006_FSHP, Bourges2017_BoSM} (see main text). Solid curve is sample C from the measurements of Fauqu\'{e} \textit{et al.}~\cite{Fauque2006_FSHP} [Fig.~1(a)]. Photo shows the YBa$_2$Cu$_3$O$_{6.67}$ single crystal used in this work.}
\end{figure}
Figs.~\ref{orthoII_raw_IFR}(a-c) and \ref{orthoVIII_raw_IFR}(a-b) show raw data before background correction, collected by the method described in Sec.~\ref{data_collection_method}. At each temperature we have calculated an average flipping ratio, the NSF counts were then divided by this number and the results are shown in Figs.~\ref{orthoII_raw_IFR}(a-c) and \ref{orthoVIII_raw_IFR}(a-b).  Our raw spin-flip data do not show the strong temperature dependences observed by Fauqu\'{e} \textit{et al.}\cite{Fauque2006_FSHP}. For example, their (011) reflection on the YBa$_2$Cu$_3$O$_{6.6}$ sample C [Fig.~1(a)] in Ref.~\onlinecite{Fauque2006_FSHP} exhibits an increase of 20\% between 300~K and 10~K. This lack of change of flipping-ratio already indicates an absence of a temperature dependent magnetic signal. 

On the other hand the intensity of the (020) nuclear reflection clearly increases as the temperature is lowered.  We can understand the $T$-dependence of all measurements within a Debye-Waller model.  Generally, the intensity of Bragg peaks increases at low temperature due to decreased thermal vibration and is described by a Debye-Waller factor.
A simple approximation for this is given by Warren \cite{Warren1990_Warr}:
\begin{equation}
I=I_0 e^{-2M},
\end{equation}
where,
\begin{equation}
2 M = \frac{3 \hbar^2 T}{m k_B T^2_D} \left[ \Phi(x)+\frac{x}{4} \right] |\textbf{G}|^2,
\end{equation}
$m$ is the unit cell mass, $T_D$ the Debye temperature, $\Phi(x)+x/4 \approx 1+x^2/36 \ldots$ and $x=T_D/T$.  Using this approximation and $T_D$=320\;K\cite{Ginsberg1994_Gins}, we obtain predictions of the temperature dependence of all the Bragg intensities [solid lines in Figs.~\ref{orthoII_raw_IFR}(a-c) and \ref{orthoVIII_raw_IFR}(a-b)] which are consistent with our data also for the weaker (010) and (011) peaks.     
 
To compare with other studies and put bounds on a putative magnetic moment appearing at lower temperature, at each temperature we have estimated the inverse flipping ratio $R^{-1}_{\textrm{meas}}$.  We find no evidence for a temperature dependence of the NSF or the SF background, see for example Fig.~\ref{bgnd_signal_comp}(c,d).  Thus, we subtract temperature-independent NSF and SF backgrounds, determined for each $\mathbf{G}$ as described in Sec.~\ref{data_collection_method} and shown in Fig.~\ref{bgnd_signal_comp} (as an example) from the data in Figs.~\ref{orthoII_raw_IFR}(a-c) and \ref{orthoVIII_raw_IFR}(a-b). No other corrections were made when we calculate $R_{\textrm{meas}}$.  Changes in $R^{-1}_{\textrm{meas}}$ for the same Bragg peak $\textbf{G}$ and sample composition should be comparable with other studies (see Eqn.\;\ref{Eqn:FR2M_perp}). Thus, the dashed lines in Figs.~\ref{orthoII_raw_IFR}(d,e) and \ref{orthoVIII_raw_IFR}(c) show schematically (see Sec.~\ref{Sec:dis} for more details) the variation of $R^{-1}_{\textrm{meas}}$ based on the results of Fauqu\'{e} \textit{et al.}\cite{Fauque2006_FSHP} for dopings measured here.  In each case, expected changes\cite{Fauque2006_FSHP} are inconsistent with our data.  Our data is consistent with a temperature-independent flipping ratio $R_{\textrm{meas}}$ equal to the instrumental flipping ratio $R_{\textrm{inst}}$. Our values of $R_{\textrm{inst}}$ are as high as 47.

We may use Eqn.~\ref{Eqn:FR2M_perp} to convert $R^{-1}_{\textrm{meas}}$ to $\left| \mathbf{M}_{\perp}(\mathbf{G}) \right|^2$ as in the previous studies\cite{Fauque2006_FSHP,Mook2008_MSFB,Bourges2011_BoSi,Sidis2013_SiBo,Mangin-Thro2015_MSWB,Mangin-Thro2017_MLSB}.  In order to do this, we first checked that our measurements of $\left| F_N (\textbf{G}) \right|^2$ are in the kinematic limit $\sigma_{\textrm{meas}} \propto \left| F_N (\textbf{G}) \right|^2$ i.e. do not require extinction corrections\cite{Squires1996_Squi} and are consistent with the structure\cite{Jorgensen1990_JVPN} of YBa$_2$Cu$_3$O$_{6+x}$. Accurate nuclear Bragg intensities $I$ can obtained by summing (integrating) over an A3 scan (a ``$\theta$-scan'' where A3=$\theta$) or making a ``$\theta-2\theta$-scan'' such that A3=$\theta$ and A4=$2\theta$. When comparing $I$ measured at different scattering angles, $\left| F_N (\textbf{G}) \right|^2$ must be multiplied by a Lorentz factor $L$ to correct for relative time spent in the diffracting position and other resolution effects. The Lorentz factor for a triple axis spectrometer can be calculated using the Cooper-Nathans \cite{Cooper1967_CoNa} or Popovici \cite{Popovici1975_Popo} method.  For the case, where the sample mosaic $\eta$ is much less than the acceptance angle of analyzer (detector) system, we find $L=1/\sin(2\theta)$ to a good approximation in agreement with analytical calculations \cite{Iizumi1973_Iizu, Pynn1975_Pynn}. To achieve this condition no collimators were placed between the sample and the detector.  The nuclear structure factor $F_N (\textbf{G})$ was calculated using the standard formula\cite{Squires1996_Squi}:
\begin{equation}
F_N(\textbf{G}) = \sum_{d} n_d b_d \exp(i \mathbf{G} \cdot \mathbf{d}) \exp \left(-B_d \frac{\left|\mathbf{Q} \right|^2}{16 \pi^2} \right),
\label{Eqn:F_N}
\end{equation}
where $\textbf{d}$ is the position of atom $d$ in the unit cell, $b_d$ is the scattering length, $n_d$ the site occupancy and $B_d$ accounts for the Debye-Waller factor.  The structure factors were calculated using data in Jorgensen \textit{et al.}~\cite{Jorgensen1990_JVPN}.  This structure assumes that the oxygen chain site O1 is randomly occupied.  The ordering of the chain oxygens\cite{Zimmermann2003_ZSFA} has little effect on structure factors of the (010) or (011) reflections.  

\begin{figure}
\includegraphics[width=0.8\linewidth]{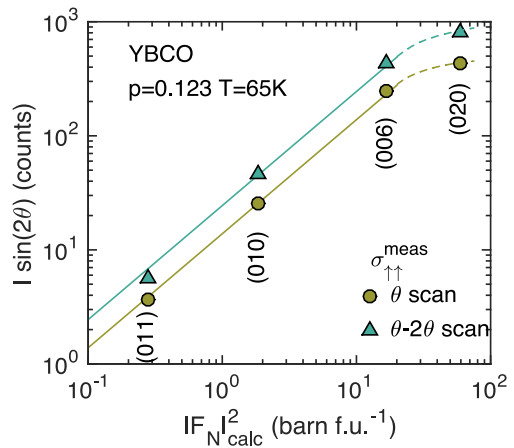}
\caption{ A comparison of the measured integrated intensity of the nuclear Bragg peaks ($p=0.123$ sample) times Lorentz correction factor with the calculated structure factors ($|F_N|^2_{\textrm{calc}}$). The data are collected under the same conditions as the rest of the experiment.  The Lorentz factor \cite{Squires1996_Squi,Iizumi1973_Iizu, Pynn1975_Pynn} corrects for different scattering angles $2 \theta$.  The structure factors are calculated using data in Jorgensen \textit{et al.}~\cite{Jorgensen1990_JVPN}. The solid lines are fits to $I \sin(2 \theta) \propto |F_N|^2_{\textrm{calc}}$. }
\label{I_vs_F_N_log}
\end{figure}

\begin{figure}
\includegraphics[width=0.7\linewidth]{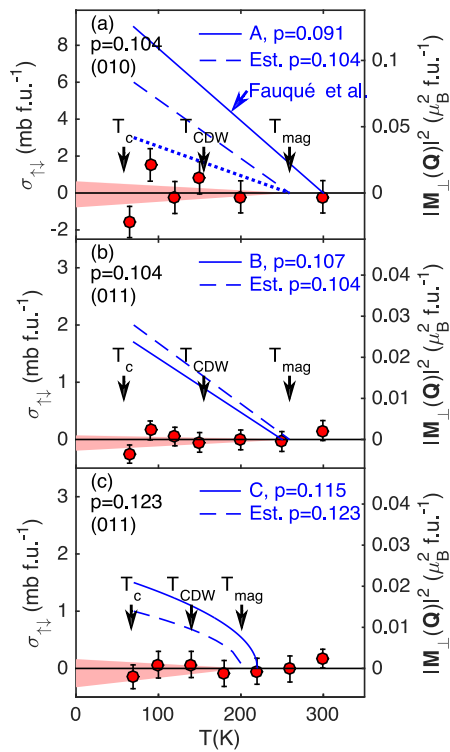}
\caption{$T$-dependence of the magnetic cross section $\sigma_{\uparrow\downarrow}$ and magnetic structure factor squared $\left|\mathbf{M}_{\perp}(\mathbf{G}) \right|^2$ determined from $R_{\textrm{meas}}^{-1}$ in Figs.~\ref{orthoII_raw_IFR}(d-e) and \ref{orthoVIII_raw_IFR}(c). Eqns.~\ref{Eqn:sigma_est} and \ref{Eqn:FR2M_perp} are used to carry out conversions with $|F_N|^2$=1.85, 0.28\;barn \;f.u.$^{-1}$ for \textbf{G}=(010) and (011). Filled pink region shows the one standard deviation range of linear fits to the data. Arrows show onset temperatures for superconductivity (Table~\ref{Tbl:samplesum}), $T_c$, 2-D charge density wave order\cite{Chang2012_CBHC,Blackburn2013_BCHH}, $T_{\textrm{CDW}}$ and putative orbital ordering\cite{Fauque2006_FSHP}, $T_{\textrm{mag}}$. The meanings of the solid, dashed and dotted lines is explained in the main text and captions to Figs.~\ref{orthoII_raw_IFR} and \ref{orthoVIII_raw_IFR}.}
\label{magmo_Tdep}
\end{figure}

Fig.~\ref{I_vs_F_N_log} (see also Table~\ref{Tbl:F_N}) shows integrated nuclear Bragg intensities with the Lorentz correction
obtained from $\theta$ and $\theta-2\theta$ scans of (A3,A4) plotted against $\left| F_N (\textbf{G})_{\textrm{calc}} \right|^2$. The data were collected under the same experimental conditions as those in Figs.~\ref{orthoII_raw_IFR} and \ref{orthoVIII_raw_IFR}. The figure shows the expected linear behavior up to the (006) reflection followed by a saturation due to extinction effects for the strong nuclear (020) reflection for both types of scan.  We find that our observed variation of the integrated intensity is consistent with the published structure of  YBCO. This verifies our normalization procedure.  In Fig.~\ref{magmo_Tdep}, we have used Eqns.~\ref{Eqn:sigma_est} and \ref{Eqn:FR2M_perp} together with the respective $\left| F_N (\textbf{G})_{\textrm{calc}} \right|^2$ to convert the data in Figs.~\ref{orthoII_raw_IFR} and ~\ref{orthoVIII_raw_IFR} into the magnetic cross section $\sigma_{\uparrow\downarrow}$ and $\left| \mathbf{M}_{\perp}(\mathbf{G}) \right|^2$.  Since there is no evidence of a $T$-dependent magnetic signal in our data, in each case we have taken $R^{-1}_{\textrm{inst}}$ to be the average value of the measured inverse flipping ratio. 

\begin{table}[b]
\begin{ruledtabular}
\begin{tabular}{l|lllll}
(hkl) &  $I\sin(2\theta)$(meas) & $|F_N|^2$(calc) & $|F_N|^2$(fit) \\
      &   (arb. units) & (barn f.u.$^{-1}$) & (barn f.u.$^{-1}$) \\ \hline
& \multicolumn{3}{c}{$\theta$ scan}\\ 
(011)  &  3.68 $\pm$ 0.12 & 0.28 & 0.27  \\ 
(010)  &  25.1 $\pm$ 0.3 & 1.85 & 1.82  \\ 
(006)  &  243 $\pm$ 5 & 16.7 & 17.5  \\ 
(020)  &  432 $\pm$ 6 & 59 &   \\ 
& \multicolumn{3}{c}{$\theta$-$2 \theta$ scan}\\ 
(011)  &  5.6 $\pm$ 0.2 & 0.28 & 0.23  \\ 
(010)  &  45.4 $\pm$ 0.7 & 1.85 & 1.86  \\ 
(006)  &  430 $\pm$ 8 & 16.7 & 17.6  \\ 
(020)  &  810 $\pm$ 13 & 59 &   \\ 
\end{tabular}
\end{ruledtabular}
\caption{\label{Tbl:F_N} Measured integrated intensity of the nuclear Bragg peaks ($p=0.123$ sample) times Lorentz correction factor compared with the calculated structure factors ($|F_N|^2$). Data from Fig.~\ref{I_vs_F_N_log}.}
\end{table}

In order to quantify our sensitivity to magnetic signal, from which we would determine the orbital magnetic moment, we fit a linear $T$-dependence $\sigma_{\uparrow\downarrow}(T)=\sigma_{\uparrow\downarrow}^{0} \times (T_{\textrm{mag}}-T)/T_{\textrm{mag}}$ to $\sigma_{\uparrow\downarrow}$ below $T_{\textrm{mag}}$ in Fig.\;\ref{magmo_Tdep}. The pink regions in Fig.\;\ref{magmo_Tdep} are bounded by the standard deviation of the fits. Table\;\ref{table:meas_moments} shows values of $\sigma_{\uparrow\downarrow}^{0}$ and the corresponding 
$\left|\mathbf{M}_{\perp}(\mathbf{G}) \right|^2$.  As mentioned above, our observations are consistent with the absence of a magnetic signal $\left|\mathbf{M}_{\perp}(\mathbf{G}) \right|$, thus our experiment simply puts an upper bound on the values of the putative orbital moment. The relationship between $\left|\mathbf{M}_{\perp}(\mathbf{G}) \right|$ and $\textbf{m}_0$ is model dependent.  To show the significance of our result we have converted our upper bound for $\left|\mathbf{M}_{\perp}(\mathbf{G}) \right|$ to corresponding upper bounds on $\textbf{m}_0$ for the various patterns in Fig.\;\ref{Fig:orb_pattern_3d} using the conversion factors in Table\;\ref{Table:M_perp_cals}. 
The results are shown in Table\;\ref{table:meas_moments}.  Previous experiments\cite{Bourges2011_BoSi} on YBa$_2$Cu$_3$O$_{6+x}$ have been analyzed in terms of the CC-$\theta_{II}$ pattern (b) and yielded $|\mathbf{m}_0|$=0.1\;$\mu_B$. Our value (Table\;\ref{table:meas_moments}) for YBa$_2$Cu$_3$O$_{6.54}$ is less than 0.013\;$\mu_B$ for this pattern. Note that each value in the lower part of Table\;\ref{Table:M_perp_cals} places a constraint on the values of $|\mathbf{m}_0|$. Thus, in the case of YBa$_2$Cu$_3$O$_{6.54}$ where two Bragg peaks are investigated, the smaller bound should be taken.

\begin{table}[t]
\begin{ruledtabular}
\begin{tabular}{m{2cm}|lll}
\multicolumn{1}{c|}{$y$ in YBCO} & 6.54 & 6.54 & 6.67 \\
\centering $(hkl)$  & (010) & (011) & (011) \\ \hline
{\centering $\sigma_{\uparrow\downarrow}^{0}(\mathbf{G})$ \\ (mbarn\;f.u.$^{-1}$)} & $-0.075 \pm 0.70$ & $-0.063 \pm 0.133$ & $-0.08 \pm 0.25$ \\
& & & \\
{\centering $\left|\mathbf{M}_{\perp}(\mathbf{G}) \right|^2$ \\ (m$\mu_B^2$\;f.u.$^{-1}$)} & $-1.0 \pm 9.8$ & $-0.9 \pm 1.9$	& $-1.2 \pm 3.4$ \\ 
& & & \\

\centering pattern & \multicolumn{3}{c}{$\left|\mathbf{m}_0 \right|$  ($\mu_B \textrm{triangle}^{-1}$)} \\ \hline  
\centering (b)  &	$<0.037$ &		$<0.013$ & $<0.020$ \\
\centering (c)	&   $<0.081$ &		$<0.09$	  &	$<0.14$ \\
\centering (d)	&   $<0.058$ &		$<0.08$	  &	$<0.12$ \\
\centering (e)	& \multicolumn{1}{c}{$-$}		 &      $<0.030$  &	$<0.046$ \\
\centering (f)	&   $<0.095$ &		$<0.09$	  &	$<0.13$ \\
\end{tabular}
\end{ruledtabular}
\caption{\label{table:meas_moments} Measured magnetic cross sections $\sigma_{\uparrow\downarrow}^{0}$ and structure factors $\left|\mathbf{M}_{\perp}(\mathbf{G}) \right|^2$ as determined from fits shown in Fig.\;\ref{magmo_Tdep}. The errors are the one standard deviation bounds determined from $\chi^2$ fitting. The one standard deviation upper bound in $\left|\mathbf{M}_{\perp}(\mathbf{G}) \right|$ is converted to an upper bound in the orbital moment per triangle ($|\textbf{m}_0|$) for the orbital patterns (b)-(f) shown in Fig.\;\ref{Fig:orb_pattern_3d} using the factors in Table\;\ref{Table:M_perp_cals}.}

\end{table}

\section{Discussion}
\label{Sec:dis}
The main finding of our experiment is that we \textit{do not} observe a temperature-dependent signal due to magnetic ordering in the two high-quality samples of YBa$_2$Cu$_3$O$_{6+x}$  that we have studied.  Although our samples are not exactly the same dopings as those studied by other groups, their compositions fit into the doping interval (see Fig.\;\ref{YBCO_phase_diagram}) where the putative order has been reported so that we could expect its presence.  In order to make an accurate comparison with the data of Fauqu\'{e} \textit{et al.}\cite{Fauque2006_FSHP} and subsequent reports\cite{Mook2008_MSFB,Bourges2011_BoSi,Sidis2013_SiBo,Mangin-Thro2015_MSWB,Mangin-Thro2017_MLSB,Bourges2017_BoSM} we estimate (see Fig.~\ref{YBCO_phase_diagram} and Table~\ref{Tbl:samplesum}) values for $\sigma_{\uparrow\downarrow} (T \approx T_c)$ of the magnetic cross section for our dopings by interpolating in between published results.  Fig.~\ref{YBCO_phase_diagram} illustrates this for the $\mathbf{G}=(011)$ reflection.  The expected temperature variation\cite{Fauque2006_FSHP} of $R_{\textrm{meas}}^{-1}$ and $\sigma_{\uparrow\downarrow}$ are shown by the blue dashed lines in Figs.~\ref{orthoII_raw_IFR},\ref{orthoVIII_raw_IFR} and \ref{magmo_Tdep}.  The continuous blue lines show the measured behavior for the samples of closest doping (labelled A, B, C) from Fauqu\'{e} \textit{et al.}\cite{Fauque2006_FSHP}.  For the $\mathbf{G}=(010)$ reflection, we compare with the twinned sample A of Fauqu\'{e} \textit{et al.} and scale the results using the $I_{\textrm{mag}}(p)$ line in Fig.~\ref{YBCO_phase_diagram} yielding the dashed lines in Figs.~\ref{orthoII_raw_IFR} and \ref{magmo_Tdep} for this reflection with $\sigma_{\uparrow\downarrow} (T \approx T_c)=6$\;mb\;f.u.$^{-1}$.  Mangin-Thro \textit{et al.} \cite{Mangin-Thro2017_MLSB} have recently claimed that there is a strong anisotropy between the magnetic signal observed at the (100) and (010) peaks, with the (010) being $\sim 3$ times weaker than (100). If this is the case, we would need to scale the dashed line by a further factor of 1/2. This is shown as the dotted lines in Figs.~\ref{orthoII_raw_IFR} and \ref{magmo_Tdep}. Note this scaling is not required\cite{Mangin-Thro2017_MLSB} for the (011) reflection.  
   
Our experiment is performed on smaller samples than the previous studies which inevitably means we have poorer statistics. However, our results still achieve the necessary statistical significance to show that the signal of magnetic order observed by Fauqu\'{e} \textit{et al.}\cite{Fauque2006_FSHP} is not present in our samples.  The signal in the neutron experiments is $\propto \left|\mathbf{m}_0 \right|^2$ rather than $\propto \left|\mathbf{m}_0 \right|$ so that the difference between the two measurements is most clearly seen in Fig.\;\ref{magmo_Tdep}.  The origin of the difference between the present and previously reported measurements is not clear, it could either be due to different physical properties of the samples or to differences in the measurement procedure. Our samples were grown by a self-flux technique using BaZrO$_3$ crucibles \cite{Erb1995_ErWF,Liang1998_LiBH}. This method of growth is known to suppress the inclusion of impurities and secondary phases in the samples, that could cause a $T$-dependent depolarization of the neutron beam or appearance of an additional magnetic signal.   The samples used by Fauqu\'{e} \textit{et al.}\cite{Fauque2006_FSHP} were grown by a melt texture growth method \cite{Marinel1997_MWMD} or top-seeded solution growth with ZrO$_2$ or Al$_2$O$_3$ crucibles\cite{Lin2002_LiLC}.  The essential differences in the present measurement procedure (see Sec.~\ref{exp_method}) are the use of samples occupying a much smaller volume and the careful re-alignment at each temperature facilitated by the high-precision mechanics of the IN20 spectrometer.  

Other probes have been used to search for magnetic order of the pseudogap phase in cuprate superconductors. It is notable that neither NMR or muon spin rotation ($\mu$SR) provide evidence for moments $\sim$$0.1\mu_B$. For example, Wu \textit{et al.}\cite{Wu2015_WMKH} discuss this explicitly. Thus our results are consistent with NMR and $\mu$SR experiments. However, the story does not end here.  The original motivation for the search for orbital magnetic order was to understand the nature of the broken symmetry in the pseudogap (PG) phase. Anomalies which may correspond to a broken symmetry have now been seen by macroscopic probes including the Kerr effect\cite{Xia2008_XSDK}, resonant ultrasound\cite{Shekhter2013_SRLH}, and optical second-harmonic generation\cite{Zhao2017_ZBLB}. The measurements have been used to construct a  phase diagram\cite{Shekhter2013_SRLH} in which the pseudogap temperature $T^{\star}$ decreases with doping.  These macroscopic probes are not directly inconsistent with our measurements since they are sensitive to more general broken symmetries or changes in the anisotropy of the system and not just magnetic order.

\section{Conclusion}
In summary, we have used polarized neutron diffraction to search for orbital magnetic order in the pseudogap phase of the underdoped high-temperature superconductor YBa$_2$Cu$_3$O$_{6+x}$. Within the sensitivity of our measurements, we do not observe such order.  This is in agreement with $\mu$SR and NMR observations and in contrast to other neutron measurements\cite{Fauque2006_FSHP,Mook2008_MSFB,Bourges2011_BoSi,Sidis2013_SiBo,Mangin-Thro2015_MSWB,Mangin-Thro2017_MLSB}, which reported signals an order of magnitude larger than the detection limit of the present experiment. During our measurements we found that, under certain circumstances, the flipping ratio measured on a sharp Bragg peak may drift significantly with temperature; this effect can be largely eliminated by sample re-alignment at each temperature. We show that the previous reported magnetic signal is not a universal (intrinsic) property of high-quality cuprate superconductor single crystals and we place a model-dependent upper bound on the magnitude of the orbital magnetic moments which is about an order of magnitude lower than the values found in previous experiments.      

\section{Ackowledgements}
We acknowledge P.\;Bourges, J.\;Chang, B.\;Fauqu\'{e}, E.\;M.\;Forgan, S.\;Lederer, S.\; A.\;Kivelson, C.\;M.\;Varma and J. A. Wilson for stimulating discussions and comments on the manuscript. The work was supported by the UK EPSRC (Grant No. EP/J015423/1).
    
%

\end{document}